# Controlling structural phases of Sn through lattice engineering

Chandima Kasun Edirisinghe[1,†], Anjali Rathore[1,†], Taegeon Lee[2], Daekwon Lee[2], An-Hsi Chen[3], Garrett Baucom[4], Eitan Hershkovitz[4], Anuradha Wijesinghe[1], Pradip Adhikari[1], Sinchul Yeom[3], Hong Seok Lee[1,2], Hyung-Kook Choi[1,2], Hyunsoo Kim[5], Mina Yoon[3], Honggyu Kim[4], Matthew Brahlek[3], Heesuk Rho[2], Joon Sue Lee[1,*]

[1]Department of Physics and Astronomy, University of Tennessee, Knoxville, TN 37996, USA

[2]Department of Physics, Research Institute Physics and Chemistry, Jeonbuk National University, Jeonju 54896, South Korea

[3]Material Science and Technology Division, Oak Ridge National Laboratory, USA

[4]Department of Materials Science & Engineering, University of Florida, Gainesville, FL 3261, USA

[5]Department of Physics, Missouri University of Science and Technology, Rolla, MO 65409, USA

[†]These authors contributed equally to this work.

*e-mail: jslee@utk.edu

## ABSTRACT

Topology and superconductivity, two distinct phenomena offer unique insight into quantum properties and their applications in quantum technologies, spintronics, and sustainable energy technologies. Tin (Sn) plays a pivotal role here as an element due to its two structural phases, α-Sn exhibiting topological characteristics and β-Sn showing superconductivity. Here, we demonstrate precise control of these phases in Sn thin films using molecular beam epitaxy with systematically varied lattice parameters of the buffer layer. The Sn films exhibit either β-Sn or α-Sn phases as the buffer layer's lattice constant varied from 6.10 Å to 6.48 Å, spanning the range from GaSb (like InAs) to InSb. The crystal structures of α- and β-Sn films are characterized by x-ray diffraction and confirmed by Raman spectroscopy and scanning transmission electron microscopy. Atomic force microscopy validates the smooth, continuous surface morphology. Electrical transport measurements further verify the phases: resistance drop near 3.7 K for β-Sn superconductivity and Shubnikov-de Haas oscillations for α-Sn topological characteristics. Density functional theory shows that α-Sn is stable under tensile strain and β-Sn under compressive strain, aligning well with experimental findings. Hence, this study introduces a platform controlling Sn phases through lattice engineering, enabling innovative applications in quantum technologies and beyond.

## 1. INTRODUCTION

Tin (Sn), a ubiquitous group IV element, finds extensive utilization in diverse industrial domains encompassing electronic circuitry, optoelectronic apparatus, energy storage systems, and surface coating applications [1−4]. Recently, interest was further triggered with the promise of exploring exotic topological phases in Sn [5−7] as well as in Sn–semiconductor hybrid materials [4,8,9]. In these novel explorations, the structural phase of Sn (α-Sn or β-Sn) plays a pivotal role. α-Sn phase has a face centered cubic (FCC) structure {Fig. 1(a)} and is a semimetal with a zero-gap protected by the cubic symmetry of its crystal lattice [3,5,10]. The application of a positive or negative in-plane biaxial strain will break this crystal symmetry and transform α-Sn to a topological insulator or a Dirac semimetal respectively [7,11−14]. Topological surface states in α-Sn have been experimentally investigated using α-Sn films under biaxial compressive strain by preparing α-Sn (bulk lattice constant: 6.49 Å) on closely lattice-matched semiconductor substrates such as InSb (lattice constant: 6.48 Å) and CdTe (lattice constant: 6.48 Å) [5,15−17]. In contrast, β-Sn is metallic, has a body-centered tetragonal (BCT) crystal structure {Fig. 1(b)} and exhibits superconductivity with a bulk critical temperature of 3.7 K [18−21]. Consequently, hybrid structures of β-Sn and low-dimensional semiconductors with strong spin-orbit coupling such as InSb or InAs [4,8,9,22,23], could potentially manifest topological superconductivity in the presence of magnetic fields [24−27].

Realizing the full potential of Sn requires understanding how to controllably grow phase-pure α-Sn and β-Sn as well as controlled phase coexistence at atomically sharp interface, a challenge particularly pronounced in the context of thin film growth because precise tuning of the phases during the growth is difficult [20,28,29]. In superconducting applications, the presence of α-Sn islands significantly reduces the yield of devices [20]. This non-uniformity gives rise to reduced performance and compromised structural stability due to the spontaneous transition between α and β phases, thereby amplifying intricacies in precise fabrication of phase-pure Sn [4].

Previously, α-Sn thin films have been grown on substrates including GaAs(001) [14], InSb(001) [14,16,30−32], InSb(111) [5], CdTe(001) [15,17,33], and Co(0001) [34] and β-Sn thin films on Si(111) [35], 6H-SiC(0001) [36], and Si(100) [37]. Additionally, Sn has been hybridized with nanowires of InAs, InAsSb and InSb [4,8,9] where problematic coexisting α and β phases have been widely observed. A comprehensive exploration into the engineering of the two phases of Sn in thin films remains to be undertaken. This knowledge is pivotal in achieving the selective growth of Sn phases and regulating the interfacial properties in Sn thin films as well as Sn-based hybrid heterostructures [4].

In this work, we demonstrate a platform that utilizes lattice engineering to control the structural phases of Sn. This is achieved by growing Sn thin films on buffer layers with varying lattice constants. We systematically varied the lattice constants by preparing buffer layers of GaSb and $In_xAl_{1-x}Sb$ with a wide range of $x$ on GaSb(001) substrates, enabling precise control over the phase selectivity of the Sn thin films. The theoretical calculations conducted also align well with the experimental results of obtaining phase pure Sn growth. The findings enrich the fundamental knowledge base in thin film technology and present a practical pathway for fabricating high-quality Sn films with phase selectivity. The advancement has the potential to influence studies of quantum phenomena in Sn films and heterostructures, as well as semiconductor technology and device manufacturing.

## 2. EXPERIMENTAL METHODS

Sn thin films were grown on buffer layers of GaSb and $In_xAl_{1-x}Sb$ on GaSb(001) substrates, as illustrated in Fig. 1(c), employing molecular beam epitaxy (MBE). To create clean GaSb (001) surface for growth 1 cm × 1 cm sized GaSb(001) substrates Ga-bonded on tantalum substrate holders and then outgassed at 300 °C for 1 hour in an ultrahigh vacuum (UHV) chamber with a base pressure of low $10^{-10}$ Torr. Subsequently, the substrates were transferred to a III-V MBE chamber, and the native oxide was removed by thermal desorption at 540 °C in the presence of $Sb_2$ flux. Then, a GaSb homoepitaxial layer of 100 nm was grown at 480°C. The resulting buffer layer exhibited at streaky (1×3) reflection high energy electron diffraction (RHEED) pattern, providing *in-situ* confirmation of the reconstructed GaSb surface achieved under Sb-rich growth conditions [38–40]. Subsequently, 700-nm-thick $In_xAl_{1-x}Sb$ heteroepitaxial buffer layers were grown. The systematic variation of the lattice constant of the buffer layers is shown as blue plus signs in (d). Eight samples were prepared with lattice constants from 6.10 Å to 6.48 Å. Buffer layers of GaSb and $In_xAl_{1-x}Sb$ were used, achieving lattice constants of 6.10 Å (GaSb), 6.20 Å ($In_{0.2}Al_{0.8}Sb$), 6.30 Å ($In_{0.48}Al_{0.52}Sb$), 6.35 Å ($In_{0.6245}Al_{0.3755}Sb$), 6.40 Å ($In_{0.77}Al_{0.23}Sb$), 6.45 Å ($In_{0.92}Al_{0.08}Sb$), 6.465 Å ($In_{0.96}Al_{0.04}Sb$), and 6.48 Å (InSb). To further fine-tune the epitaxial growth and minimize defects between the substrate and the subsequent Sn layer, the growth temperatures of the buffer layers were varied between 450°C and 350°C. This strategic approach aimed to ensure a crystalline growth of epitaxial films and enhance the overall structural integrity of the system [41,42]. Finally, the substrates were transferred to an interconnected UHV chamber without breaking the vacuum, and 30-nm-thick Sn films were grown on top of the buffers. To maintain optimal conditions during the Sn growth, active cooling with liquid nitrogen was employed, ensuring that the process occurred well below room temperature preventing segregation of Sn blobs due to its high surface mobility as well as any unwanted phase transition from α-Sn to β-Sn initiated due to high temperature [43].

No spontaneous phase transition from α-Sn to β-Sn is expected in the 30-nm-thick Sn films. Although bulk Sn exhibits the structural phase transition at 13°C, Sn films at reduced thicknesses show the transition at higher temperatures [29]. For example, 8-nm-thick α-Sn thin films grown on InSb are reported to be stable above room temperature up to around 180°C [44]. An α-Sn film with a thickness of 60 nm shows the transition occurring at 115°C [45]. Hence, the 30-nm-thick Sn films in this work are likely to be stable during the growth procedure. Right after the Sn film growth, samples were transferred to a load lock within a few minutes, and the load lock was vented with $O_2$ gas to oxidize the top layers of Sn. This oxide layer ensures that any change in the surface morphology is prevented, and the Sn thin film is thermodynamically stable [46].

Crystal structures of Sn films and the buffer layers were characterized by x-ray diffraction (XRD). Further strain/relaxation of the Sn films were characterized by reciprocal space mapping (RSM). XRD and RSM were measured using Malvern Panalytical X'pert3 Materials Research Diffractometer with Cu-kα radiation source (λ=1.54 Å). The atomic structure of a strained α-Sn film was characterized by high-angle annular dark-field (HAADF) imaging in scanning transmission electron microscopy (STEM). Cross sectional lamella was prepared using a FEI Helios Nanolab 600i Dual Beam focused ion beam scanning electron microscopy with a final polishing acceleration voltage of 2 kV. HAADF-STEM imaging was performed using an aberration corrected Themis Z (Thermo Fisher) microscope with an acceleration voltage of 200 kV, a probe semi convergence angle of 22 mrad and a 50 – 200 mrad collection angle. Surface morphology of the grown Sn films was characterized by atomic force microscopy (AFM) using a Cypher S AFM machine. Two measurements were carried out for each sample covering a wider area of 10 × 10 μm² and a smaller area of 1 × 1 μm².

For Raman mapping measurements, an excitation argon-ion laser light with a wavelength of 514.5 nm was used. The excitation laser power was maintained at less than 1 mW at which laser-induced heating of the sample was not observed and the measurements were done at room temperature. Magneto-transport measurements on all the grown thin films were done with magnetic field up to 14 T at 2 K in a Quantum Design Dynacool Physical Property Measurement System. For this measurement, Hall bars were created by scratching the top surface of the Sn samples and placing Indium dots on the Hall bar patterns, with rough dimensions of 0.5 mm in width and 1.3 mm in length.

## 3. RESULTS AND DISCUSSION

### A. Structural characterization by XRD

The lattice constant of the buffer layers and their influence on the crystallography of the Sn phases were characterized with $\theta/2\theta$ XRD measurements on eight Sn samples, as shown in Fig. 1(e). Each scan exhibits peaks due to the α- or β-Sn phases, as well as the buffer layer and substrate, as indicated. The key finding of this data is that the β-Sn phase is observed when the lattice constant of the buffer layer falls below 6.35 Å, whereas the α-Sn phase is observed when the lattice constant is 6.35 Å or greater. Furthermore, both α-Sn(001) and β-Sn(001) thin films were epitaxially matched to the buffer layers of $In_xAl_{1-x}Sb(001)$.

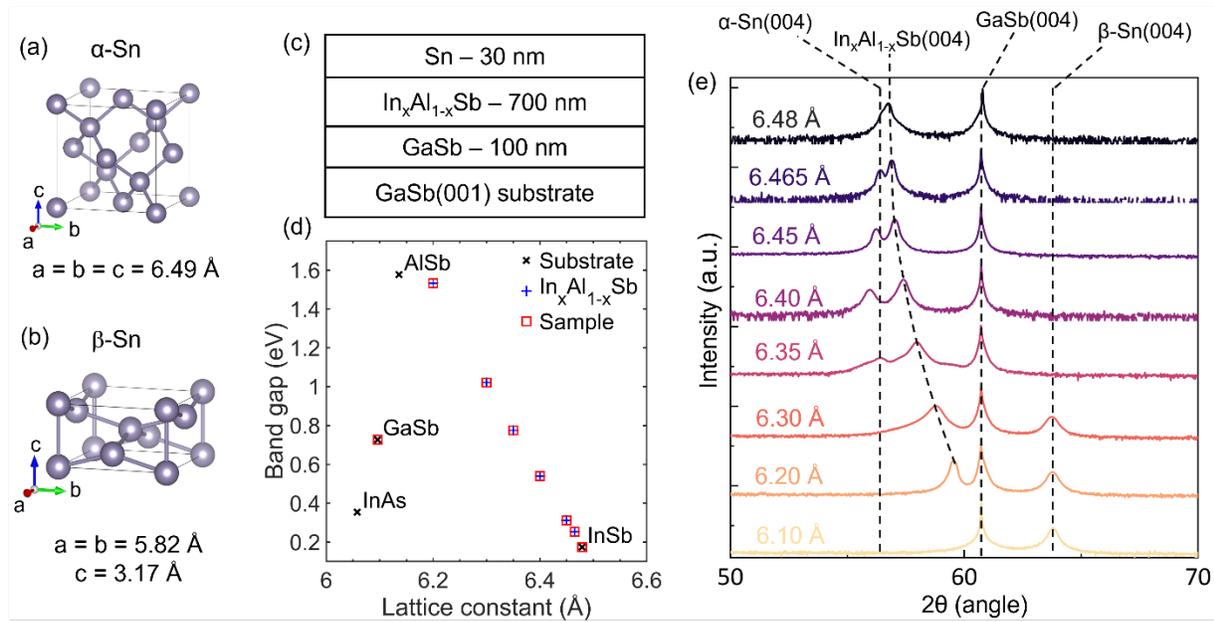

Figure 1. Crystalline study of Sn thin films. (a) FCC crystal structure of α-Sn. (b) BCT crystal structure of β-Sn. The lattice constants of the lattice structures are denoted below the crystal structures in (a) and (b). (c) Schematic of the sample stacking of Sn thin films grown on GaSb(001) substrate with the $In_xAl_{1-x}Sb$ buffer layer. (d) Band gap as a function of lattice constant for different substrates and the buffer layers used in this study. (e) Stacked $\theta/2\theta$ measurement results for the samples with the lattice constant of the buffer layer, GaSb and $In_xAl_{1-x}Sb$, systematically varying from 6.10 Å (bottom) to 6.48 Å (top).

Out of plane lattice constants were calculated using the α-Sn(004) XRD peaks, and strain was calculated utilizing the bulk lattice parameter of the α-Sn, focusing on samples with lattice constants exceeding 6.35 Å, as shown in Table 1. As the obtained lattice constant of α-Sn(004) peak increases from 6.40 Å to 6.48 Å, a notable decrease in the calculated out-of-plane uniaxial tensile strain is observed. The sample featuring a lattice constant of 6.35 Å shows a drop in the strain value within the α-Sn phase, indicating relaxation of α-Sn layers. The reason for this phenomenon is likely stemming from the heavy strain imparted on the α-Sn layer due to the pronounced lattice mismatch of 2.16% between the buffer layer and the α-Sn thin film. It might be highly strained α-Sn at the interface, which gradually relaxes towards the surface with increasing thickness. In a highly strained film, as film thickness increases, more defects such as misfit dislocations tend to form to relieve a portion of the strain due to the lattice mismatch, allowing the epitaxial layer to relax towards its strain-free lattice constant [47]. We note that the calculated strain value might slightly differ from the actual value because the Poisson ratio for epitaxial mismatch and the out-of-plane lattice parameter were not considered [48].

Table 1: Out-of-plane strain on α-Sn films calculated from XRD results.

| Nominal lattice constant of buffer | Lattice mismatch between buffer and α-Sn | Lattice constant of α-Sn from(004) peak | Strain on α-Sn calculated from obtained lattice constant |
| --- | --- | --- | --- |
| 6.48 Å | 0.15 % | 6.51 Å | 0.35 % |
| 6.465 Å | 0.39 % | 6.52 Å | 0.45 % |
| 6.45 Å | 0.62 % | 6.53 Å | 0.74% |
| 6.40 Å | 1.39 % | 6.56 Å | 1.18 % |
| 6.35 Å | 2.16 % | 6.52 Å | 0.47 % |

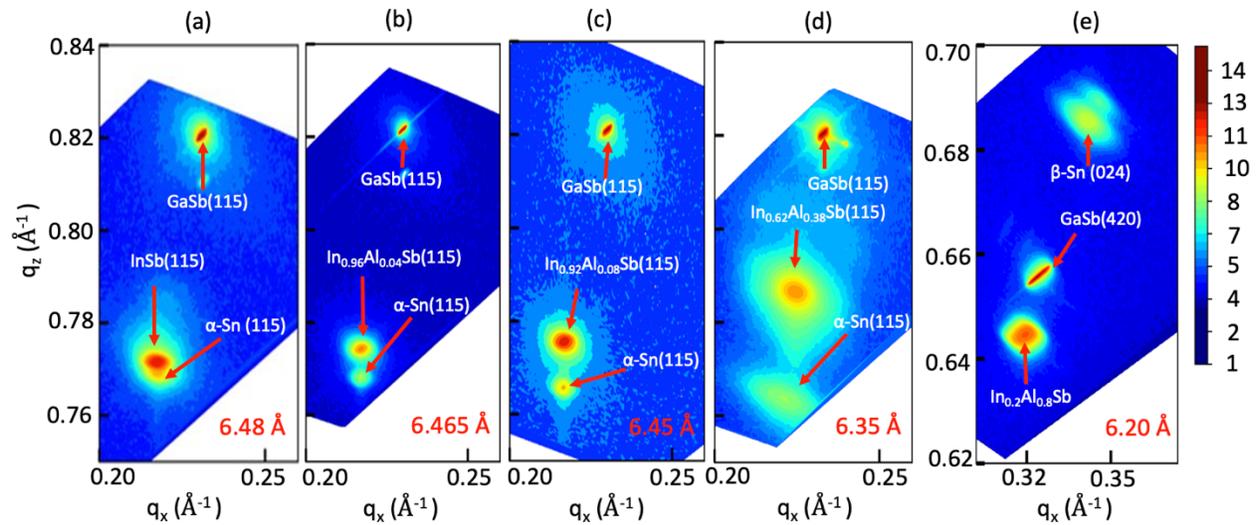

Figure 2. RSM scans of samples with different buffer layer lattice constants. RSM images for samples of (a) 6.48 Å, (b) 6.465 Å, (c) 6.45 Å, and (d) 6.35 Å with α-Sn phase show strained thin films except for 6.35 Å sample, (e) RSM shows relaxed thin films for β-Sn phase in sample 6.20 Å.

To comprehensively understand the strain on Sn layers, reciprocal space mapping (RSM) was systematically conducted on Sn thin films with different lattice constants of buffer layers, as shown in Figure 2. For the α-Sn film grown on InSb buffer layer with a lattice parameter of 6.48 Å, as shown in Fig. 2(a), the presence of strain was observed, as indicated by the α-Sn(115) peak and InSb(115) sharing the same $q_x$. For samples with a lower lattice parameter, a clear separation between the α-Sn peak and the buffer layer peak was noted {Figs. 2(a-c)}. There is also an increase in strain on the α-Sn thin films, consistent with our findings from XRD peak positions (Table 1). Importantly, the sample with a lattice parameter of 6.35 Å was completely relaxed {Fig. 2(d)} compared to those with higher lattice parameters, supporting the conclusions drawn from the $\theta/2\theta$ measurements. Conversely, relaxed β-Sn(024)/GaSb(420) peaks were detected for samples with lattice parameters of 6.20 Å, as shown in Fig. 2(e) with small compressive strain of 0.19% calculated from lattice parameter obtained from β-Sn(004) peak.

## B. Surface morphology by AFM

Surface morphology of the Sn samples was investigated by utilizing AFM, as shown in Fig. 3. All the Sn films show continuous and smooth surfaces. The surface roughness can be quantified by root mean square (RMS) roughness, and the RMS roughness of all the Sn films turned out to be less than 2 nm from 1 μm × 1 μm scans as indicated in Fig. 3(d). Strained α-Sn films {Fig. 3(a)} are smoother than relaxed α-Sn {Fig. 3(b)) and relaxed β-Sn films {Fig. 3(c)) which is revealed in Fig. 3(d) with higher RMS roughness in Sn films with the buffer lattice constant below 6.40 Å. To attain a more refined understanding of the surface morphology, additional AFM scans were performed across a larger area of 10 μm × 10 μm, as shown in Fig. S2 of the Supplemental Material, which align with the trend observed for the 1 μm × 1 μm scans, with strained α-Sn films {Fig. S2(a)} exhibiting smoother surfaces compared to both relaxed α-Sn {Fig. S2(b)} and relaxed β-Sn films {Fig. S2(c)}.

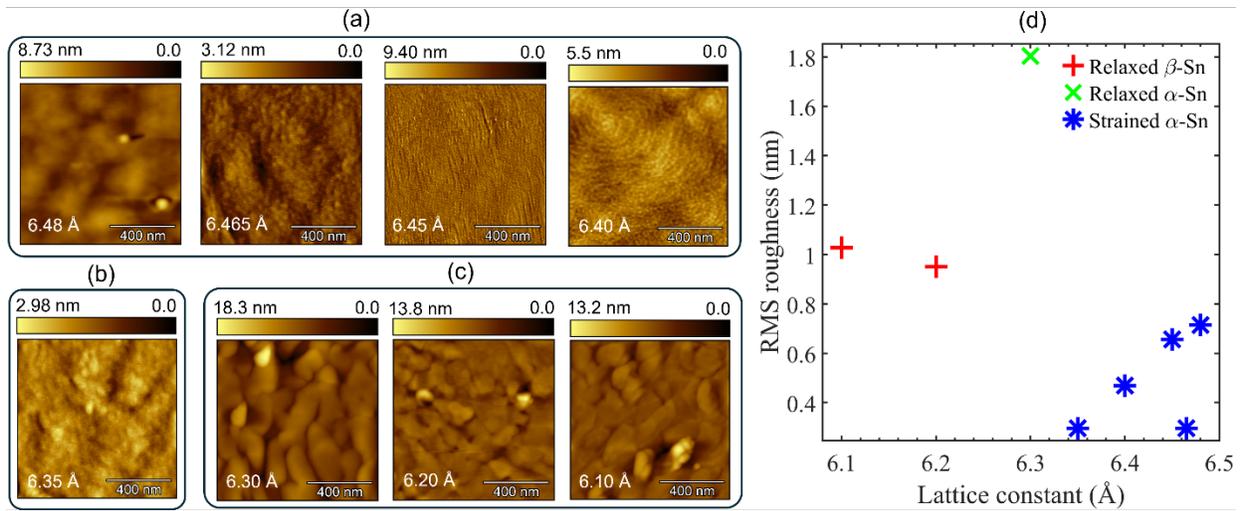

Figure 3. AFM images depicting the surface morphology of (a) strained α-Sn, (b) relaxed α-Sn, and (c) relaxed β-Sn phases of the samples. (d) RMS roughness of the samples as a function of buffer layer lattice constant, for a surface area of 1 μm × 1 μm.

It is particularly noteworthy that the RMS roughness values are higher for samples with lattice constants of 6.30 Å and 6.35 Å, as shown in Fig. S2(d). These samples exhibit grainy features with larger heights, which can be attributed to the structural transition occurring between the β- and α-phases of Sn within the lattice constant range spanning from 6.30 Å to 6.35 Å. This transition region represents a critical juncture where the relaxed α-Sn phase emerges, notably observed at the lattice constant of 6.35 Å. Importantly, it is in this transition region where the lattice mismatch between the buffer layer and the α-Sn or β-Sn layer reaches its peak. This increased lattice mismatch is directly correlated with an increased occurrence of defects within the thin film. Consequently, the presence of defects will manifest as grainy features with large heights in the surface morphology of the Sn films. In Fig. 3(d), this phenomenon is not readily apparent due to the limited analysis area which does not capture the full extent of the surface features and variations present in the samples. Hence, observations derived from AFM underscore the intricate interplay between lattice mismatch, phase transitions, and surface morphology in Sn thin films.

## C. Distinctive Raman response in α- and β-Sn films

To obtain distinct spectral fingerprints characteristic of both the α- and β-Sn phases, thus offering additional validation of the phase engineering of Sn, Raman scans were meticulously conducted, as delineated in Fig. 4. Raman spectra acquired from Sn films grown on two distinct buffer layers of $In_xAl_{1-x}Sb$ which corresponded to lattice constants of 6.48 Å and 6.20 Å exhibit distinctive features indicative of α- and β-Sn phases, respectively.

For the Sn film grown on the InSb buffer layer with a lattice constant of 6.48 Å, a prominent Raman peak at 197 cm$^{-1}$ is observed, which corresponds to the optical phonon mode of single crystalline α-Sn [31]. A weaker Raman response at 187 cm$^{-1}$, appearing as a low-frequency shoulder of the α-Sn phonon peak, is attributed to the longitudinal optical (LO) phonon mode of InSb, [49,50] indicating the presence of the InSb buffer layer. In addition to the mentioned peaks, weak Raman responses at 116 cm$^{-1}$ and 150 cm$^{-1}$ are discerned, which correspond to the optical phonons of antimony (Sb) [51]. This observation suggests the presence of segregated Sb below the α-Sn layer, likely formed on the InSb buffer layer during the cool down process right after the InSb buffer layer growth with the Sb shutter opened, which was then closed below 300 °C substrate temperature.

Additionally, within the acquired Raman spectra, distinct responses are identified within the frequency regions proximal to 230 cm$^{-1}$ and within the range of 360–390 cm$^{-1}$. Notably, the broad mode centered around 230 cm$^{-1}$ can be attributed to the presence of the SbSn phase [52], a proposition reinforced by evidence from XRD data presented in Fig. S1. The peak at 41.2° corresponds to either the SbSn or the $Sb_3Sn_4$ phase. The ambiguity between SbSn and $Sb_3Sn_4$ arises due to their identical crystal structures [53]. The broad Raman response observed within the 360–390 cm$^{-1}$ frequency region is attributed to multi-phonon scattering phenomena, primarily stemming from interactions involving both InSb and α-Sn phases [31,50].

For the Sn film grown on the $In_xAl_{1-x}Sb$ buffer layer with a lattice constant of 6.20 Å, of particular interest is the absence of the robust 197 cm$^{-1}$ mode associated with α-Sn in the Raman spectrum of the Sn film grown on the $In_xAl_{1-x}Sb$ buffer layer. Instead, a novel Raman peak emerges at 125 cm$^{-1}$, a characteristic feature not observed for α-Sn. The 125 cm$^{-1}$ mode corresponds to the optical phonon of single crystalline β-Sn [31,54]. The absence of the 197 cm$^{-1}$ mode indicative of the α-Sn phase fraction suggests that the Sn crystal grown on $In_xAl_{1-x}Sb$ lacks the α phase fraction. Analogous to the observation in α-Sn, weak Raman responses attributed to Sb are also discerned at 116 cm$^{-1}$ and 150 cm$^{-1}$ within the Raman spectrum of the β-Sn phase. It is noteworthy that the broad 230 cm$^{-1}$ mode is not observed in β-Sn. The broad response in the 360–390 cm$^{-1}$ region is not also detectible.

The unequivocal findings from the Raman spectroscopic analysis underscore the remarkable capability to systematically manipulate the structural phase of Sn by finely tuning the lattice constant of the $In_xAl_{1-x}Sb$ buffer layer.

To investigate the structural uniformity of the α-Sn phase, spatially resolved Raman mapping measurements were conducted, yielding valuable insights into variations in peak frequencies and spectral widths of the optical phonon of α-Sn across the investigated area of 10 × 10 µm$^2$. The spatial distribution of peak frequencies extracted from the α-Sn optical phonon is shown in Fig. 4(b). The Raman image depicting variations in the spectral widths of the α-Sn optical phonon is presented in Fig. 4(c). The spatially averaged values of the frequencies and widths are $\bar{\omega} = 196.7 \pm 0.1$ cm$^{-1}$ and $\bar{\Gamma} = 5.9 \pm 0.1$ cm$^{-1}$, respectively, indicative of the structural uniformity and crystalline coherence of the α-Sn film.

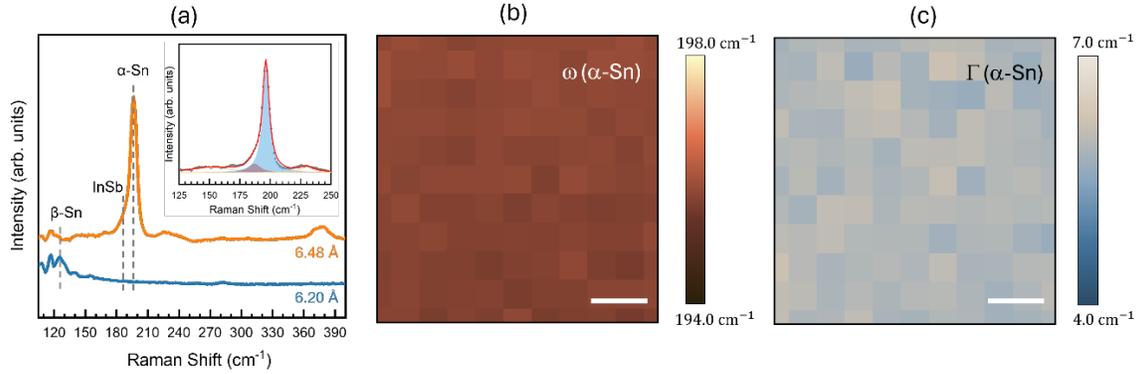

Figure 4. (a) Representative Raman spectra of α-Sn (top, 6.48 Å) and β-Sn (bottom, 6.20 Å). The inset shows a fitted result of the Raman spectrum of α-Sn, showing deconvoluted individual components of the 150, 187, 197, and 230 cm$^{-1}$ modes. Spatial variations in (b) peak frequencies and (c) spectral widths of the optical phonon of α-Sn. The scale bars are 2 µm.

### D. Epitaxial α-Sn film observed by STEM

A cross-sectional view of the 6.45 Å sample with atomic resolution is presented in Fig. 5(a), revealing the epitaxial nature of both the buffer layer In$_{0.92}$Al$_{0.08}$Sb and the α-Sn layer. Fig. 5(a) is significant considering the lattice mismatch of 0.62%, as shown in Table 1. Because of this lattice mismatch, defects emerge at the interface and in the grown layers. These imperfections are observed in the form of misfit dislocations and twinning defects, as showcased in Figs. 5(b) and 5(c), respectively. The burgers circuit is drawn in orange around the dislocation core identifying the Burgers vector as $\vec{b} = \frac{1}{2}\langle 110 \rangle$.

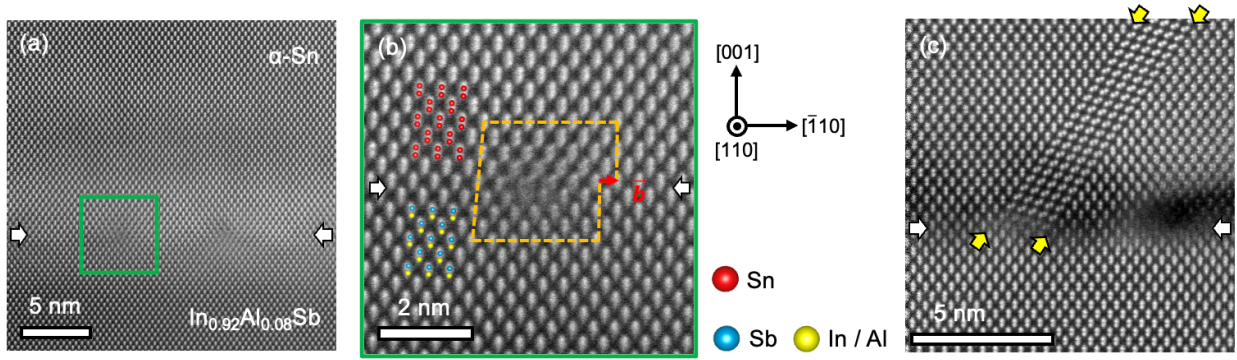

Figure 5. STEM images of the 6.45 Å sample. (a) HAADF-STEM image of the In$_{0.92}$Al$_{0.08}$Sb and α-Sn interface. (b) Magnified HAADF-STEM image of the green boxed region in (a) highlighting a misfit dislocation. (c) A twining defect at the interface spans across the α-Sn film. The twin boundaries are highlighted by yellow arrows. In all three images, the interface between In$_{0.92}$Al$_{0.08}$Sb and α-Sn is marked with white arrows.

### E. Electrical transport in α- and β-Sn films

To understand the topological and superconducting properties, electrical characterization of the both α-Sn and β-Sn thin films as a function of temperature and external magnetic field conducted, as depicted in Fig. 6. Superconductivity was observed in samples exhibiting the β-Sn phase. Figs. 6(a) and 6(b) resistance measurements for a representative β-Sn film with a lattice constant of 6.2 Å, with varying temperature and perpendicular magnetic field. A sharp superconducting transition occurred at 3.8 K with an abrupt resistance drop down to zero resistance, which is consistent with bulk β-Sn superconducting transition [54–56]. The resistance drop becomes less pronounced and disappears at a magnetic field of 0.25 T. The transition temperature was corroborated by the plot of longitudinal resistance versus magnetic field, shown in Fig. 6(b), where the first resistance drop flattened at 3.8 K. In addition to the transition at 3.8 K, two other resistance drops at around 4.7 K and 5.8 K were observed. Similar phenomena were noted in the investigation by Ding et al. [57], in which they explored the possibility of localized superconducting transitions within β-Sn islands and α-Sn mediated island coupling. However, unlike in our study, the presence of β-Sn islands and an α-Sn phase was not observed in this sample. Therefore, further research is required to thoroughly comprehend the underlying factors contributing to the observation of multiple resistance drops.

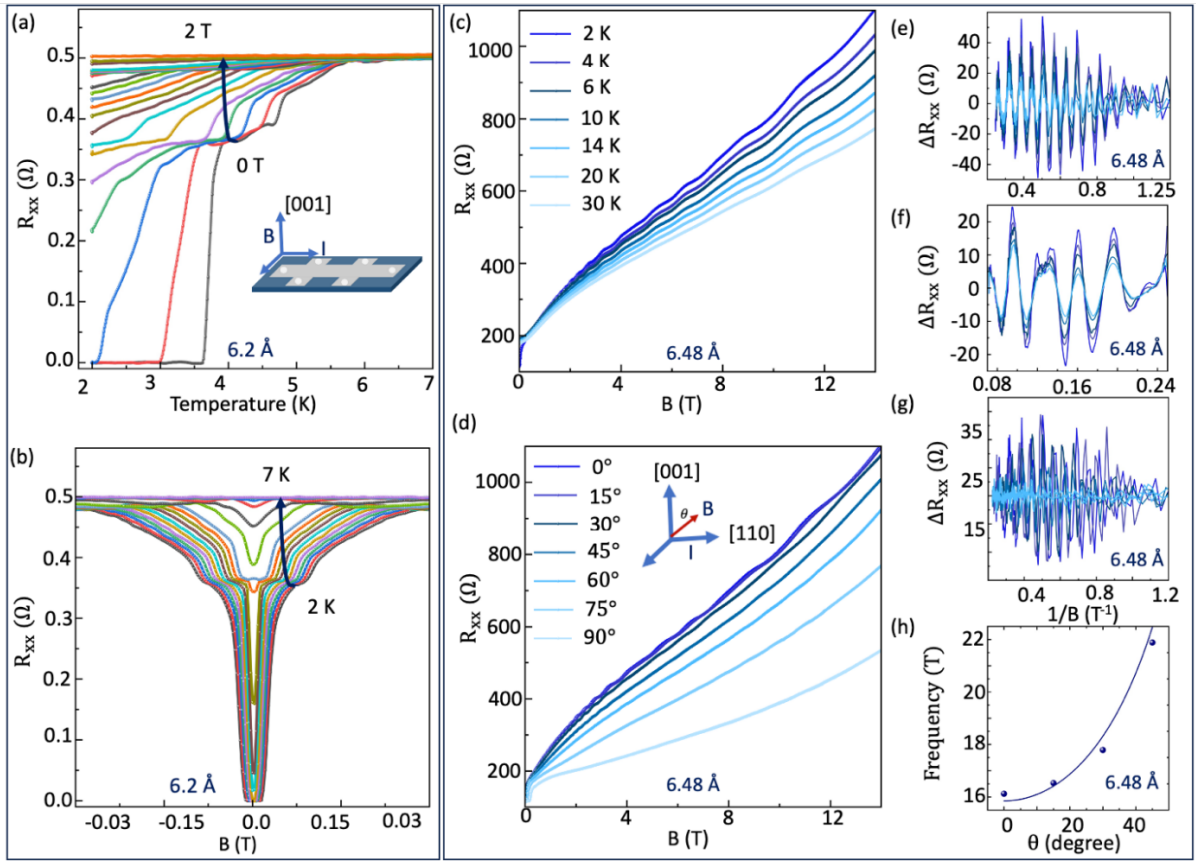

Figure 6. Magneto-transport measurements for both the 6.20 Å and 6.48 Å samples. (a) $R_{xx}$ versus temperature plot with varying magnetic field for the 6.20 Å sample, (b) $R_{xx}$ versus magnetic field plot with varying temperature for the 6.20 Å sample, (c) Depicts the SdH oscillations for $R_{xx}$ versus B under an applied perpendicular magnetic field with varying temperatures, (d) SdH oscillations for $R_{xx}$ versus B with varying angles of magnetic field, (e) and (f) present the $\Delta R_{xx}$ versus 1/B with varying temperatures for both the low and high magnetic field ranges, respectively, (g) Shows the $\Delta R_{xx}$ versus 1/B for the low magnetic field range with varying angles, (h) show the FFT frequency versus θ with 1/cos(θ) fitting.

Magneto-transport measurements on an α-Sn film with lattice constant of 6.48 Å showed clear Shubnikov-de Haas (SdH) oscillations starting from 0.8 T. Temperature dependence of the SdH oscillations in $R_{xx}$ with applied perpendicular magnetic field shows a weakening trend of the amplitude of the oscillations with increasing temperature and disappearance of the oscillations beyond 20 K {Fig. 6(c)}. Based on the observed periodicity in 1/B of oscillations within the magnetic field range of 0-14 T, distinct oscillations occur in the low field range spanning 0.8 - 4 T and a high field range spanning 4 - 14 T. Figs. 6(e) and 6(f) present background corrected $R_{xx}$ versus 1/B at various temperatures for the high and low magnetic field ranges, respectively. Oscillation frequencies were determined via Fast Fourier transform (FFT) analysis at different temperatures, revealing a single oscillation frequency of approximately 16 T in the low magnetic field range with effective mass of $0.03m_e$, while a frequency of 30 T was observed in the high magnetic field range with effective mass of $0.07m_e$. Background correction and fitting for the calculation of effective masses are discussed in Supplemental Material. Additionally, Lifshitz-Kosevich (LK) fitting was studied on 1/B periodic oscillations for low magnetic field range, and mobility of about $1.5 \times 10^4$ cm$^2$/Vs was obtained (see Supplemental Material S6).

As the frequence of the SdH oscillations probes the extremal cross section of the Fermi surface, field dependence of the frequence vs angle of the field enables exploring the Fermi surface geometry. Variations in the magnetic field direction relative to the current direction at different angles θ are shown in Fig. 6(d). Here, 0° represents an orientation with the magnetic field normal to the current and surface and 90° denotes a parallel alignment to the current direction with the field in the plane of the sample. Notably, clear angle dependent SdH oscillations were evident with varying the angles up to 45°, while from 60° to 90°, the oscillations became broader and eventually disappeared as shown in Figs. 6(g). The FFT frequencies of these oscillations were determined within both the low and high magnetic field ranges and plotted against θ as depicted in Fig. 6(h). To understand the angular dependence, an ellipsoidal Fermi surface will follow a cos(θ) dependence whereas a 2D Fermi surface follows a 1/cos(θ) dependence since the Fermi surface is open normal to the plane. As such, the data were fitted with a 1/cos(θ) function, represented by a solid line in the plot. The cos(θ) dependence of the oscillation frequency was particularly notable in the low magnetic field range, indicative of the possible presence of two-dimensional (2D) carriers.

Given that this sample was grown on an InSb buffer layer, the potential for parallel conduction arises. To examine this, individual electrical transport measurements were performed on InSb buffer layers grown on GaSb substrates, which showed no oscillations (see Supplemental Material Fig. S7). Consequently, it can be inferred that the observed oscillations originate from the α-Sn.

The α-Sn phase is known to alter its band structure in response to strain. Moreover, previous studies have indicated that the α-Sn phase can exhibit different topological phases under in-plane compressive or tensile strain. [7,16] Notably, prior investigations of α-Sn film grown on an InSb substrate have demonstrated the existence of topological surface states (TSS) between the second valence band ($\Gamma^{7-}$ s band) and the conduction band ($\Gamma^{8+}$ p band) through angle-resolved photoemission spectroscopy measurements. [16,58] Furthermore, several studies have observed the existence of SdH oscillations in α-Sn films [59] and have concluded these oscillations originate from the TSS. [17,60] Thus, for the α-Sn film in this work, it is likely that the lower field oscillations originate from the TSS characterized by low effective mass carriers (0.03 $m_e$) with high mobility (1.5 $\times 10^4$ cm$^2$/Vs). This contrasts with the oscillations observed in the high magnetic field range, which likely emanate from bulk states of heavy holes, illustrating a three-dimensional (3D) nature.

## F. Theoretical investigation of stability of α- and β-Sn films

To explore the stability of α- and β-Sn films under various strains induced by different lattice constants, a theoretical investigation was carried out by obtaining the α-Sn and β-Sn structures from the Materials Project [61] and relaxing them fully using the Vienna Ab initio Simulation Package (VASP) [62,63]. The relaxed lattice parameters for α-Sn and β-Sn are $6.64 \times 6.64 \times 6.64$ Å³ and $5.91 \times 5.91 \times 3.25$ Å³, respectively. The following steps were performed to determine the relaxed structure at specific strains along the a/b axis. First, supercells of $1 \times 1 \times 2$ α-Sn and $1 \times 1 \times 3$ β-Sn were prepared. Then, a given strain was applied to the a/b axis of each supercell while keeping the total volume fixed. The a/b strained structures were then used to generate a set of structures strained along the c axis only, without changing the a/b lengths. The atomic positions of these structures were relaxed, their energies were obtained, and the minimum energy and corresponding c length were found from a fitted curve of energy versus c. The minimum energy from the fitted curve was considered to be the energy at a given a/b strained lattice parameter. The relaxation conditions were as follows: The Perdew-Burke-Ernzerhof (PBE) [64] exchange-correlation functional was used, with the default energy cutoff for tin set to 103 eV. The minimum allowed distance between k-points was 0.15 Å⁻¹ with spin-orbit coupling. The conjugate gradient algorithm was used for ion relaxations, ensuring that the maximum force component was less than $5 \times 10^{-3}$ eV/Å.

Because the strain induced is biaxial in the in-plane direction, both a and b lattice parameters were changed simultaneously and the structures for the given a and b parameters were relaxed so that c is optimized for the given in-plane strains. From the results shown in Fig. 7, α-Sn is more stable than β-Sn in the pristine structures. This indicates that there is a structural transition from α-Sn to β-Sn with a critical compressive strain of 6%, above which β-Sn becomes more stable. The critical lattice constant that induces such a transition is ~6.22 Å. In other words, starting from the β-Sn structure, tensile strain would induce a structural transformation to α-Sn when the lattice constant greater than the critical point of 6.22 Å.

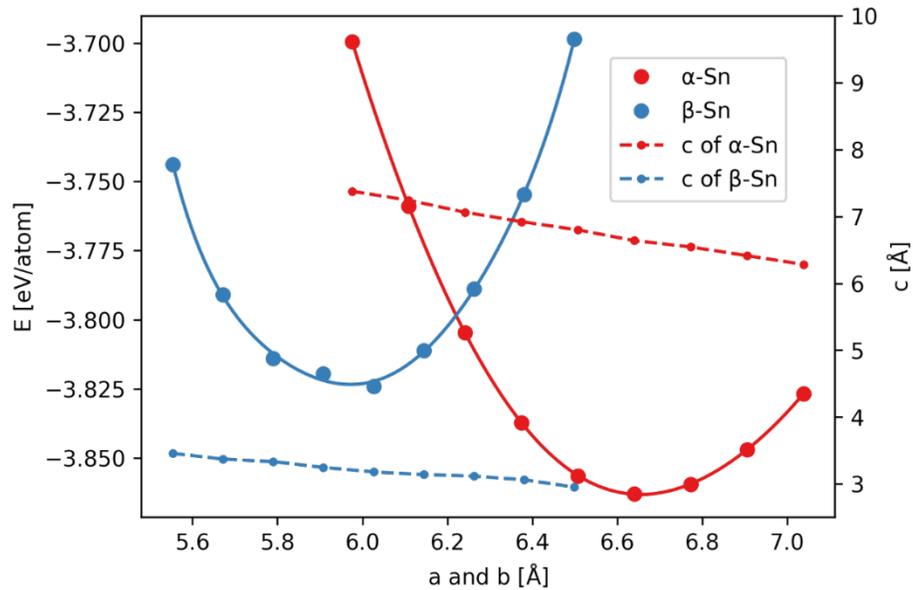

Figure 7. Total energies of α-Sn and β-Sn with different lattice parameters, *a* and *b*. The lattice parameter *c* was optimized for the given *a* and *b*. Here the total energy is divided by the number of atoms in the unit cell considered in the calculations.

These theoretical results are in good agreement with experimental observations of Sn films grown on buffer layers with different lattice constants. From the experiment, β-Sn phase is observed when the lattice constant of the buffer layer falls below 6.35 Å, whereas the α-Sn phase is observed when the lattice constant is 6.35 Å or higher. This experimental result supports the theoretical prediction that α-Sn becomes more stable than β-Sn at larger lattice parameters. The theoretical results thus provide a clear explanation for the experimental observations. The stability of one Sn phase over the other across specific lattice constant ranges confirms the experimental observation of phase pure growth of either α- or β-Sn by lattice engineering.

## 4. CONCLUSIONS

In summary, we have demonstrated epitaxial engineering of structural phases of Sn thin films by varying the lattice constant of the buffer layers and performed an in-depth characterization of the resulting phases of Sn. Sn thin films were grown on GaSb and $In_xAl_{1-x}Sb$ buffer layers on GaSb(001) substrates by MBE. A comprehensive structural analysis using XRD reveals the existence of the two different Sn phases (α-Sn and β-Sn) depending on the lattice constant of the $In_xAl_{1-x}Sb$ buffer layer. AFM results reveal higher RMS roughness near the boundary of α-Sn and β-Sn transition with varying the lattice constant. Raman spectroscopy confirms the α-Sn and β-Sn films by revealing distinct optical phonon modes, and the Raman mapping measurement reveals the uniformity and crystalline coherence of α-Sn film. In agreement with these analyses, low-temperature electrical measurements on the Sn thin films demonstrated a sharp superconducting transition for β-Sn phase and quantum oscillations for α-Sn phase which are indicative of the electronic band structures of TSS with a 2D nature and bulk heavy holes with a 3D nature. The experimental observation of the phase pure growth of Sn at different lattice constants is in good agreement with the theoretical calculations conducted. In brief, along with the optimal growth conditions, choosing the right lattice constant for the buffer layer is the key to realizing the desired Sn phases. Therefore, this work presents a crucial development to achieve a selective Sn phase in thin film growth as well as in device fabrication.

## ACKNOWLEDGMENTS

This work was supported by the Science Alliance at the University of Tennessee, Knoxville, through the Support for Affiliated Research Teams program, and the U.S. DOE, Office of Science, Basic Energy Sciences, Materials Science and Engineering Division (structural and transport characterization), by the U.S. Department of Energy, Office of Science, Office of Basic Energy Sciences, Materials Sciences and Engineering Division (S.Y.), and by the U.S. Department of Energy (DOE), Office of Science, National Quantum Information Science Research Centers, Quantum Science Center (M.Y.) and this research used resources of the Oak Ridge Leadership Computing Facility at the Oak Ridge National Laboratory, which is supported by the Office of Science of the U.S. Department of Energy under Contract No. DE-AC05-00OR22725 and resources of the National Energy Research Scientific Computing Center, a DOE Office of Science User Facility supported by the Office of Science of the U.S. Department of Energy under Contract No. DE-AC02-05CH11231 using NERSC award BES-ERCAP0024568.